\documentclass[a4paper]{article}

\usepackage{xcolor}

\usepackage[margin=1in]{geometry} 

\usepackage{amsmath}
\usepackage{amsthm}
\usepackage{amssymb}

\usepackage[utf8]{inputenc}
\usepackage{hyperref}
\hypersetup{
	unicode,
	pdfauthor={Author One, Author Two, Author Three},
	pdftitle={A simple article template},
	pdfsubject={A simple article template},
	pdfkeywords={article, template, simple},
	pdfproducer={LaTeX},
	pdfcreator={pdflatex}
}
\usepackage{soul}

\usepackage[sort&compress,numbers,square]{natbib}

\theoremstyle{plain}

\theoremstyle{definition}

\usepackage{graphicx, color}
\graphicspath{{fig/}}

\usepackage{algorithm, algpseudocode} 
\usepackage{mathrsfs} 

\usepackage{lipsum}

\usepackage{ulem}

\title{Scaling the tail beat frequency and swimming speed in underwater undulatory swimming
}
\author{Jesús Sánchez-Rodríguez$^{1,2}$, Christophe Raufaste$^{1,3}$ and Médéric Argentina$^1$}

\date{
	$^{1}$Université Côte d'Azur, CNRS, INPHYNI, France\\
$^{2}$ Laboratory of Fluid Mechanics and Instabilities, École Polytechnique Fédérale de Lausanne, CH-1015 Lausanne, Switzerland\\
$^{3}$Institut Universitaire de France (IUF), France%
}

\begin{document}
	\maketitle
	
\begin{abstract}
Due to its great efficiency and maneuverability, undulatory swimming is the predominant form of locomotion in aquatic vertebrates. A myriad of animals of different species and sizes oscillate their bodies to propel themselves in aquatic environments with swimming speed scaling as the product of the animal length by the oscillation frequency. Although frequency tuning is the primary means by which a swimmer selects its speed, there is no consensus on the mechanisms involved. In this article, we propose scaling laws for undulatory swimmers that relate oscillation frequency to length by taking into account both the biological characteristics of the muscles and the interaction of the moving swimmer with its environment. Results are supported by an extensive literature review including approximately 1200 individuals of different species, sizes and swimming environments. We highlight a crossover in length around 0.5-1 m. Below this value, the frequency can be tuned between 2-20 Hz due to biological constraints and the interplay between slow and fast muscles. Above this value, the fluid-swimmer interaction must be taken into account and the frequency is inversely proportional to the length of the animal. This approach predicts a maximum swimming speed around 5-10 m.s$^{-1}$ for large swimmers, consistent with the threshold to prevent bubble cavitation. 
\end{abstract}

\section{Introduction}

Beyond a few centimeters in length, most aquatic vertebrates propel themselves through the water by deforming their spines and propagating deformation waves through the body  \cite{lauder2005hydrodynamics}. Fish, cetaceans, reptiles, amphibians, and birds oscillate the head, body, tail, and/or fins, as appropriate, with a variety of gaits described by a specific classification \cite{sfakiotakis1999review}. Despite the complexity of treating each case separately, the kinematics of underwater undulatory swimmers can be captured to first order with a few parameters such as the wavelength of the deformation $\lambda$, the tail beat amplitude $A$ and the tail beat frequency $f$. There is now considerable evidence that $\lambda$ and $A$ are strongly related to animal length $L$, regardless of the size and shape of the animal, or the swimming conditions. In the example of fish, the wavelength scales as the animal length, with a factor of the order of unity \cite{videler_fish_1993, DiSanto2021}. The same applies to the tail beat amplitude that follows $A\simeq 0.2 L$ \cite{bainbridge1958speed, rohr2004strouhal,hunter1971swimming, saadat2017rules} from tadpoles of a few centimeters to whales of 20 meters in length (Methods \ref{Methods:Data}). These simple allometric scaling relations reveal general physical laws that are valid over several orders of magnitude in size. Momentum balance and minimal energy expenditure associated with the hydrodynamic interaction between the moving body and the surrounding water appear to drive the selection of the amplitude, as well as the determination of the swimming speed, $U \simeq 3.3 A f$ \cite{bainbridge1958speed, triantafyllou1993optimal,gazzola2014scaling, saadat2017rules}. Therefore, the swimming speed is proportional to the oscillation frequency for a given swimmer and scales as $L f$, with a proportionality factor between 0.4 and 1 for fish and cetaceans \cite{videler_fish_1991,videler_fish_1993,videler_differences_1985,curren_designs_1992,svendsen_maximum_2016}. Taking the average factor, the relationship $U\simeq0.7 L f$ is therefore a very good approximation of the swimming speed to within a factor of 2 at most. The swimming speed is thus intrinsically linked to the oscillation frequency, but unlike the aforementioned scaling laws that follow clear and widely documented trends over several orders of magnitude in length, no consensus has been reached on the law that sets the tail beat frequency. Most studies agree that the frequency decreases with the length  \cite{bainbridge1958speed} and  scaling laws $f \sim L^{-n}$ with an exponent $n$ ranging between 0.5 and 1 are often reported together with models referring to biological constraints, to the hydrodynamic interactions of the swimmer with its environment or even to the effect of gravity  \cite{bejan2006unifying,sato2007stroke,watanabe_slowest_2012,bale2014energy,gough2019scaling}. The difficulty of establishing a clear law and identifying the mechanisms at play is related to the fact there is only a factor of 100 between the highest frequency recorded in the smallest fish and the lowest frequency recorded in the largest cetaceans, typically 20 and 0.2 Hz respectively (Fig. \ref{fig:FrequencyVsLength}), while measurements show a large dispersion for a given length. Firstly, several parameters such as the previous training or the condition of the animal, its age or sex, and the water or body temperature influence the final performance of the swimmer \cite{zhao2012effects, allen2006effects, makiguchi2017sex}. Secondly, the frequency is not fixed for a given swimmer but is the parameter that is adjusted to determine the swimming speed, and a factor of 10 in frequency can easily be observed for a given specimen \cite{bainbridge1958speed}.

Obtaining experimental results that maintain homogeneity in all these features for a wide range of aquatic animal sizes is simply impossible and it is thus not surprising to find diverse experimental laws in the literature. In particular, comparing animals with the same level of activity, such as sustained, prolonged or burst, would be necessary \cite{wu1977introduction} but difficult to implement in experiments \cite{brett1964respiratory,brett_metabolic_1972}. As a consequence, instead of focusing on a specific activity gait, we propose to gather all data available in the literature, regardless of the level of activity or any other specific characteristic, to build a database of more than one thousand entries.  
This approach provides a complete picture of the dependency of the frequency with the length and captures both the main trend and the dispersion associated with a given length. 
We thus propose a frequency selection mechanism that, on the one hand, balances the swimmer's muscle force and the reactive forces generated by the fluid when the animal is in motion and, on the other hand, considers the type of muscles, slow and fast. This works uncovers allometric relations for the swimming frequency and speed that take the form of scaling laws in the limits of very small or very large swimmers.

\section{Results}

\subsection{Tail beat frequency measurements}

\begin{figure}[H]
    \centering\includegraphics[width=\textwidth]{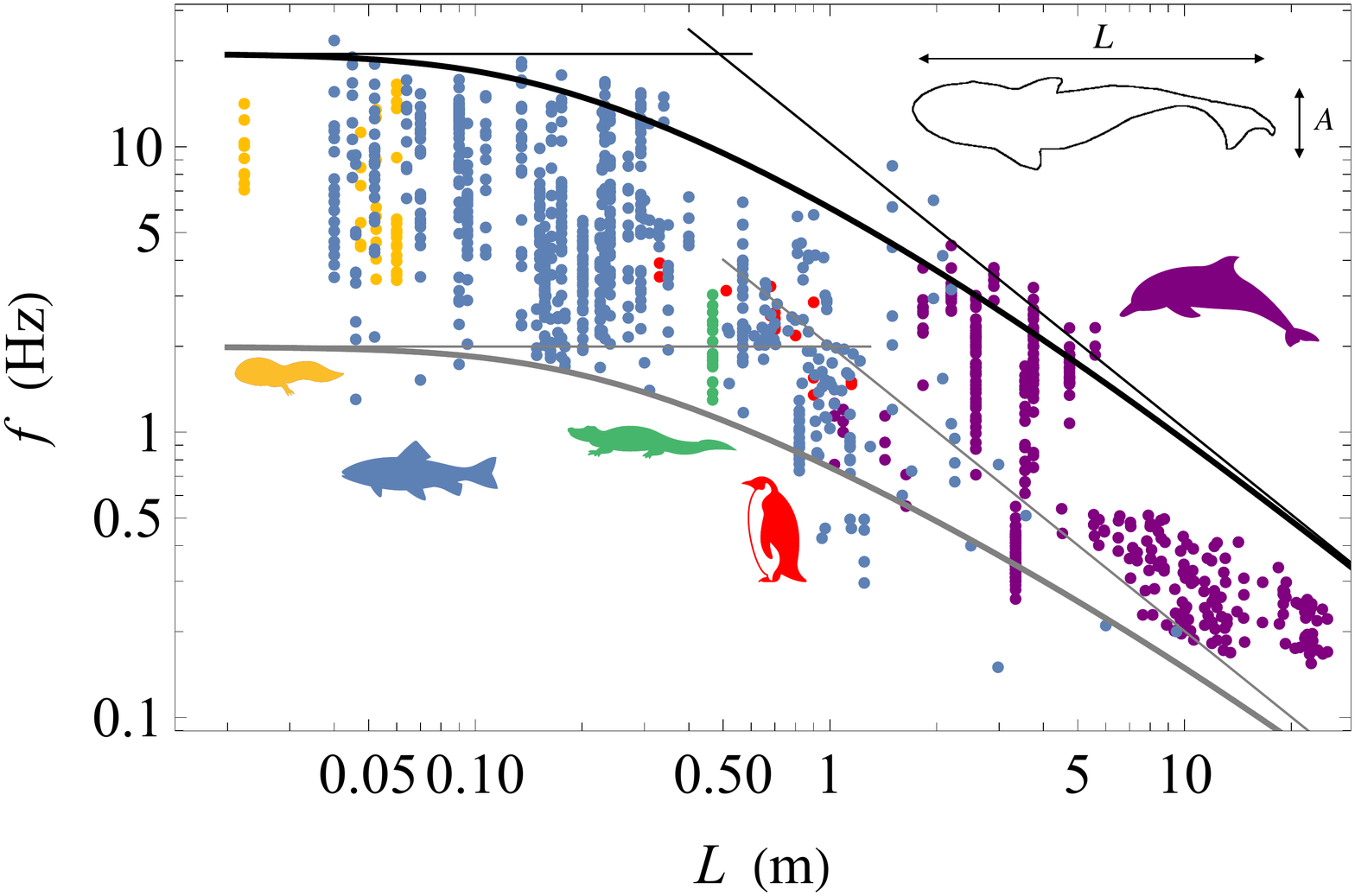}
    \caption{Tail beat frequency $f$ as a function of length $L$ for amphibians (yellow), fish (blue), reptiles (green), birds (red) and mammals (purple). Thick black and grey lines represent the burst and sustained activity levels, respectively, fitted with the model. Thin lines are the scaling laws in the limits of very small and very large swimmers. Values of the parameters for the fast bound are $\left[L_c\right]_\mathrm{fast} = (0.5 \pm 0.2)$ m, $\left[f_0\right]_\mathrm{fast} = (21\pm 2 )$ Hz, $\left[\kappa\right]_\mathrm{fast} =(4 \pm 3)$, while for the slow bound we find $[L_c]_\mathrm{slow} = (1.0 \pm 0.5 )$ m, $[f_0]_\mathrm{slow}= (2.0 \pm 0.3)$ Hz, $[\kappa]_\mathrm{slow}= (9 \pm 11 )$.
    }    \label{fig:FrequencyVsLength}
    \end{figure}

We collected about 1200 data points from references listed in Methods \ref{Methods:Data}, with no discrimination on the basis of activity level or any of the other parameters mentioned above, to avoid any possible bias in the length-frequency relationship. 
In Fig. \ref{fig:FrequencyVsLength}, it appears that the tail beat frequency is correlated to the length, with all measurements located within a band in the $L-f$ plane. For most of the lengths, the upper bound is well identified as the burst activity level \cite{videler_fast_1983, fish_comparative_nodate, marras2015not} and the frequency varies approximately by a factor of 10 for a given $L$. For the longest animals, typically cetaceans with $L>5$ m, the magnitude of the band decreases. As we will discuss later, this decrease is most likely associated with a lack of burst frequency measurements: unlike the smaller animals, these animals were only observed in their natural environment and were not forced to swim at peak activity. If we define the upper and lower bounds of the band as fast and slow, we observe that both follow the same behavior: the frequency is constant and maximum at small lengths, typically $[f]_\mathrm{fast} \simeq 20$ Hz  and $[f]_\mathrm{slow} \simeq 2$ Hz respectively, before decreasing at larger $L$. The change in tendency occurs around $L = 0.5 - 1$ m. Since the span in $f$ observed for a given $L$ is of the same order of magnitude as the span in $f$ of a given specimen to adjust its speed, we conclude that the frequency intervals are primarily associated with variations in the level of activity. 

In the next section, we interpret the data set by taking into account both the muscle activity and the mechanical interaction of the swimmer with his environment. Building on previous models, we account for limitations of muscle in terms of both maximal force and frequency to explain the transition around $L = 0.5 - 1$ m between a regime characterized by a frequency that is constant and a regime where the frequency decreases with length. Finally, we consider the nature of the muscles, slow versus fast, to predict the slow and fast bounds of the band in the $L-f$ plane.

\subsection{Model and scaling laws}

Locomotion occurs as a result of undulating movements produced by the contraction of blocks of muscle segments \cite{shadwick2006fish}. While the muscles contract on one side of the swimmer, those on the opposite side relax, alternately flexing the entire body from side to side. Vertebrate muscles share a large number of structural and functional features  \cite{videler_fish_1993} that can be reasonably well described by a limited number of parameters. This is the case for the relationship between the force in the muscle,  $F$, and the muscle contraction velocity, $v$, as described by Hill's muscle model \cite{hill1938heat}:
\begin{equation}
    F = F_0 \frac{1- \displaystyle\frac{v}{v_0}  }{1+ \displaystyle \kappa \displaystyle\frac{v}{v_0} },
    \label{HillEquation}
\end{equation}
where $F_0$ is the maximum isometric force generated in the muscle, obtained as  $v$ tends to zero, and $v_0$ is the maximum contraction velocity over which no force can be produced. In its dimensionless form, the force $F/F_0$ is a decreasing and convex function of the velocity $v/v_0$, whose degree of curvature is quantified by the parameter $\kappa$ (Methods \ref{Methods:Hill}). Since the muscle fibers work in parallel, the maximum force scales as the cross area of the muscle: $F_0 \sim L^2 \sigma_0$ where  $\sigma_0$ is the maximum isometric force per unit cross-sectional area \cite{hill_dimensions_1950}. The contraction velocity $v$ drives the tail velocity and we can expect the scaling $v \sim A f$, which leads to $v/v_0=f/f_0$, where $f_0$ is the maximum tail beat frequency expected from the physiological limit of the muscle. $f_0$ can be inferred by measuring the twitch contraction time $T$ of the muscle \cite{wardle_limit_1975}, that is, the period of a single contraction and relaxation cycle produced by an action potential within the muscle fibers, in order to deduce the maximum frequency in the form $f_0 = \frac{1}{2T}$ since one period consists of two antagonistic contractions. 

The swimmer's movements set the surrounding fluid in motion to propel it along. Given that the only force exerted by the animals on the fluid is provided by the muscles, there must exist a balance between the forces produced by the muscles and the reaction force of the fluid. Above a few centimeters in length, aquatic organisms have a mode of locomotion based on inertia  \cite{gazzola2014scaling}: while the body is oscillating, boluses of water of mass $\rho L^3$ are set in motion with an acceleration $A f^2$ normal to the tail, resulting in a lateral force  that scales as $\rho L^3 A f^2$, where $\rho$ is the density of water. As discussed in the Introduction and in the Method \ref{Methods:Data}, $A$ is proportional to $L$ and thus the lateral force that pushes the fluid scales as $\rho L^4 f^2$. By balancing the latter with the force exerted by the muscle (Eq. (\ref{HillEquation})), we obtain:
\begin{equation}
    \left(\frac{L}{{L_c}}\right)^2  = \left(\frac{{f_0}}{f}\right)^2 \frac{1- \displaystyle\frac{f}{f_0}  }{1+ \displaystyle\kappa \displaystyle\frac{f}{f_0} },
    \label{eqBalance}
\end{equation}
where we have introduced the length $L_c$  that marks the crossover between two regimes with different scaling laws:
\begin{eqnarray}
    \label{eq:asymptiticsLc}
    L_c & = & \sqrt{\frac{\sigma_0}{\rho}} \frac{1}{f_0}\\
    \label{eq:asymptiticsfa}
    f &\simeq& f_0, \quad  \mathrm{if}\quad L \ll L_c\\
    \label{eq:asymptiticsfb}
    f &\simeq& f_0\frac{L_c}{L}, \quad  \mathrm{if}\quad L \gg L_c.
\end{eqnarray}
Eq. (\ref{eqBalance}) predicts nicely the two bounds of the frequency band. In Fig. \ref{fig:FrequencyVsLength}, we have drawn the best fits (Methods \ref{Methods:FitActivityRegimes}) with the set of parameters $[L_c]_\mathrm{fast} = 0.5 \pm 0.2$ m, $[f_0]_\mathrm{fast} = 21 \pm 2$ Hz and $[\kappa]_\mathrm{fast}=4 \pm 3$ for the fast bound and $[L_c]_\mathrm{slow} = 1.0\pm0.5$ m, $[f_0]_\mathrm{slow}= 2.0\pm0.3$ Hz and $[\kappa]_\mathrm{slow}=9 \pm 11$ for the slow bound. Unlike $f_0$ and $L_c$, $\kappa$ has large standard deviations from the fitted values; in fact $\kappa$ weakly shapes the fit since it only plays a role in the transition between the two limit regimes. These fits are also predictive: on the example of humans and underwater undulatory swimming, also called dolphin kick, the fit of the fast bound predicts a maximum kicking frequency around 3-4 Hz for a swimmer of about 2 meters, in good agreement with data recorded in elite swimmers \cite{matsuura_muscle_2020}.

In addition, this model reconciles the two approaches and validates each in its own length range.

For $L \ll L_c$, the frequency is fixed by a biological constraint, $f_0$, in the spirit of the approach initiated by Wardle \cite{wardle_limit_1975} who proposed that the maximum tail beat frequency corresponds to the maximum frequency expected from the muscles. 

For $L \gg L_c$, the frequency decreases with length as a consequence of the interplay between another biological constraint, here $\sigma_0$, and the interaction of the swimmer with its environment  \cite{hill_dimensions_1950,sato2007stroke,bale2014energy}. In this limit, the model predicts that the frequency is given by $f = c L^{-1}$, where we have introduced the speed $c =L_c f_0= \sqrt{\sigma_0/\rho}$, that we estimate to be $[c]_\mathrm{slow}\simeq 2.0\ \mathrm{m.s^{-1}}$ and $[c]_\mathrm{fast} \simeq 10\ \mathrm{m.s^{-1}}$ for the slow and fast bounds, respectively. The ratio of the wavelength of deformation $\lambda$ to the body length ranges from 0.74 to 1.14 for anguillorm and thunniform swimmers, respectively \cite{videler_fish_1993,DiSanto2021}, and $c\simeq \lambda f$ can also be interpreted as the speed of the wave propagating along the swimmer's body. Still in the perspective of a mechanical approach, we remark here that the scaling $f \sim L^{-1}$ from Eq. (\ref{eq:asymptiticsfb}) is also consistent with an approximation of the swimmer by an elastic beam of length $L$, radius $R$ and Young Modulus $E$.
In such a case, the natural frequency of bending waves scales as $f_e\sim \frac{R}{L^2}\sqrt\frac{E}{\rho}$ \cite{landau1986Elasticity}. By assuming geometrical similarity, $R\sim L$, and body elasticity compatible with stresses generated by muscle fibers $E\sim \sigma_0$, the natural bending frequency scales as the tail beat frequency $f$ of large animals, defined in  Eq. (\ref{eq:asymptiticsfb}). This suggests that large animals undulate their bodies to resonate with their natural bending modes, as suggested by various studies  \cite{gazzola2015gait,paraz2016thrust,hoover2018swimming}.

Finally, we now understand why there are so many different exponents in the literature resulting from attempts to describe the frequency-length relationship as a scaling law (e.g., in \cite{bejan2006unifying,sato2007stroke, watanabe_slowest_2012, bale2014energy, gough2019scaling}). Special attention must be paid to the analysis of a range of lengths, because 1) the scaling laws are only valid in either of the two limiting regimes and 2) the data must be measured at the same activity level for the comparison to be meaningful.

\subsection{Considering the type of muscle fiber and comparison with biological data}

In order to connect the biology of swimmers to the fast and slow bounds in Fig. \ref{fig:FrequencyVsLength}, we take into account the different types of muscle fibers. Muscle fibers can be roughly characterized as fast or slow, in part because the latter type has a much lower level of ATP activity and a smaller contraction speed but increased activity of oxidative enzymes. Therefore, slow fibers are intended to produce forces over a prolonged period of activity \cite{Fitts1991111}, while fast fibers use anaerobic chemical reactions and are adapted to rapid movements, while producing higher forces. As a result, fast and slow muscles are primarily solicited in burst and sustained activity levels, respectively \cite{rome1988animals}. In what follows, we hypothesize that the presence of the lower frequency boundary of the $f-L$ graph reflects the use of slow fibers, while to reach high frequencies, fast fibers are exploited. The fast and slow frequency bounds in Fig. \ref{fig:FrequencyVsLength} will therefore be modeled with the same Eq. (\ref{eqBalance}), but with parameters values that account for fast and slow muscles, respectively.

The parameters obtained from the fits match the biological data very well. Wardle et al. suggests inferring $f_0$ from measurements of the twitch contraction time \cite{wardle_limit_1975}, resulting in values between 5 and 25 Hz for 4 cm to 2.3 m fish  \cite{videler_fish_1991}. These measurements are in agreement with the value $[f_0]_\mathrm{fast} = 21\ \mathrm{Hz}$ that fits the fast bound. Given that $\sigma_0  \simeq 200$ kN.m$^{-2}$ for fast muscles  is rather constant among species \cite{johnston_thermal_1984}, the estimate $L_c \sim 0.7$ m is in excellent agreement with the fit $[L_c]_\mathrm{fast}= 0.5 \ \mathrm{m}$. For slow muscles, the maximum frequency is smaller, and we employ the same approach to study the slow bound, although measurements are rarer in this case. In the example of 10 cm salmon and 1 m sharks, $f_0$ ranges from 0.5--2 Hz, again in good agreement with the $[f_0]_\mathrm{slow}=2.0\ \mathrm{Hz}$ obtained from the fit of the slow bound. Measurements of $\sigma_0$ for slow muscles are found between 20 and 80 kN.m$^{-2}$ \cite{altringham_pca-tension_1982,johnston_thermal_1984, johnston1985force,matsuura_muscle_2020}. If we take 50 kN.m$^{-2}$ as the typical value, we find $L_c \sim 3.5$ m, whose order of magnitude is coherent with the value $[L_c]_\mathrm{slow}= 1 \ \mathrm{m}$ found from the fit. Finally, the values of $\kappa$ adjusted for the fast and slow bounds, $[\kappa]_\mathrm{fast}=4$ and $[\kappa]_\mathrm{slow}=9$, are in the  range of values recorded in vertebrates, typically between 2.5 and 10  \cite{wilkie1949relation, johnston1985sustained}.

This framework with six parameters, three for each bound, appears coherent because the filament arrangement in striated muscles is very similar along all vertebrates  \cite{hooper2005invertebrate}. 

In addition to the activity level, temperature has an effect on the twitch contraction time, and therefore on $f_0$  \cite{wardle_limit_1975, wardle_effects_1980}. Cold water generally depresses locomotory muscle function: the lower the temperature, the lower $f_0$ is and a factor of 5 difference can be easily found between 2 and 30$^\circ$C. This can lead to deviations from the main trend at extreme temperatures. This is probably the main explanation of the surprising low tail beat frequency of Greenland sharks in Arctic waters ($L \simeq 3$ m and $f\simeq 0.15$ Hz in Fig. \ref{fig:FrequencyVsLength}), a factor of 2--3 below the fit of the slow bound \cite{watanabe_slowest_2012}. Conversely, there are examples of thermal acclimation of fish in warm waters that could lead to $f_0$ values as high as 50 Hz \cite{johnson_thermal_1995}.

\subsection{Scaling the swimming speed}

The relationship $U \simeq 0.7 L f$ intrinsically relates the tail beat frequency to the swimming speed to a very good approximation, with a factor of 2 at most for fish and cetaceans for $U$. This allows us to infer scaling laws for a given activity level:
\begin{eqnarray}
    \label{eq:asymptiticsUa}
    U &\simeq& 0.7 f_0 L, \quad  \mathrm{if}\quad L \ll L_c\\
    \label{eq:asymptiticsUb}
    U &\simeq& 0.7 f_0 L_c, \quad  \mathrm{if}\quad L \gg L_c.
\end{eqnarray}
For small animals, we therefore expect swimming speed to increase with length, whereas it should saturate at a constant value, $0.7 \sqrt{\sigma_0/\rho}$ from Eqs. (\ref{eq:asymptiticsLc}) and (\ref{eq:asymptiticsUa}), for large animals. This is consistent with studies that found a tail beat frequency scaling as $L^{-1}$ for large swimmers, and also found a nearly constant swimming speed in this case \cite{sato2010scaling, bale2014energy}. Still for large swimmers, the swimming speed should range approximately between 1 and 8 m.s$^{-1}$ depending on the activity level, given the values of $f_0$ and $L_c$ found from the fits of the slow and fast bounds, respectively.

\begin{figure}[htb]
    \centering\includegraphics[width=\textwidth]{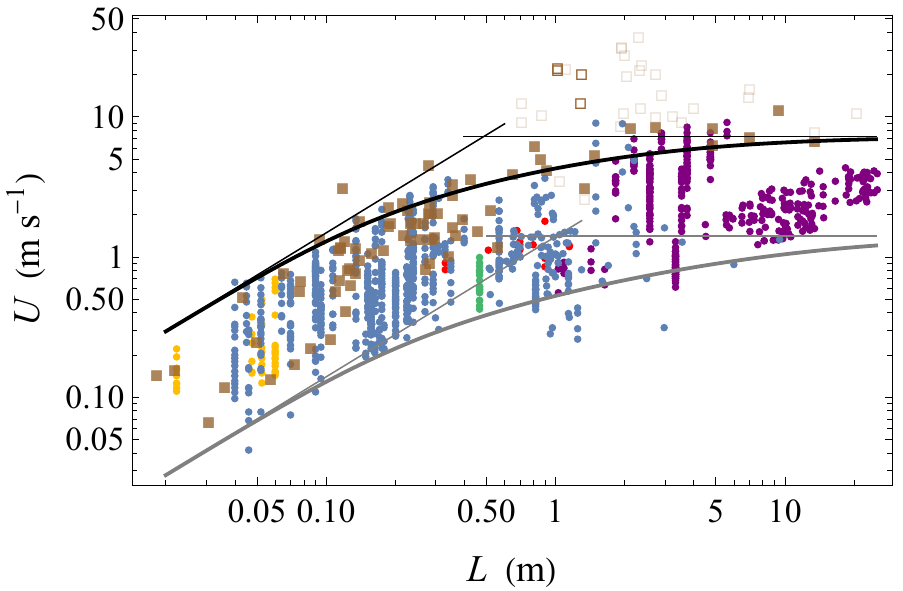}
    \caption{
    Swimming speed $U$ as a function of length $L$. Following the law  $U=0.7 L f$, we show our estimates of swimming speed from the tail beat frequency measurements displayed in Fig. \ref{fig:FrequencyVsLength} (closed circles): amphibians (yellow), fish (blue), reptiles (green), birds (red), and mammals (purple). 
    Brown squares correspond to the data gathered by Hirt et al. for maximum swimming speeds \cite{hirt2017general} using the mass-length relationship (Methods \ref{Methods:Data}). 
    Open translucent squares represent either non-peer reviewed papers or data coming from estimations and not actual measurements. 
    Open opaque squares represent data obtained using rod-mounted devices. The other data are represented by closed opaque squares.    
    The black and gray thick lines represent the fast and slow boundaries predicted by the model and fitted parameters in Fig. \ref{fig:FrequencyVsLength}, respectively. Thin lines are the scaling laws in the limit of very small and very large swimmers.}  
    \label{fig:VelocityVsLength}
    \end{figure}

In order to properly describe the mechanisms at play, it is essential to compare data at the same activity level. 
We therefore rely on the study by Hirt et al.,  who reported data on maximum swimming speed \cite{hirt2017general}. In Fig. \ref{fig:VelocityVsLength}, we superimpose these data on the expected swimming speeds inferred from all frequency measurements. Various points need to be discussed.

First for $L \lesssim 0.5 - 1$ m, we observe that both data sets exhibit the same upper limit. For most lengths, the match is perfect and only small differences are observed, but they are at most a factor of 2. This regime is consistent with the scaling law expected in the limit $L < L_c$: quantitatively, Hirt et al. found that the maximum swimming speed scales as $U \propto M^{0.36}$, on average, or equivalently $U \propto  L$, in agreement with the scaling law proposed in Eq.  (\ref{eq:asymptiticsfa}). 

Second, for $L \gtrsim 0.5 - 1$ m, we observe a significant difference. While our scaling law inferred from frequency measurements predicts a constant speed, around 5--10 m.s$^{-1}$, data gathered by Hirt et al. suggest a humped shape with a maximum around 30--40 m.s$^{-1}$ obtained for $L \simeq 1 - 3$ m followed by smaller speeds for larger animals. 
In fact, we propose that the two data sets differ in the two regions for two different reasons.

\begin{enumerate}
\item
While Hirt et al. did an enormous amount of work gathering data from the literature, we suggest that the maximum is artificial if we apply relevant filters. Most of the highest maximum speed data collected by Hirt et al., for fish ranging from about 1 to 3 m in length, are estimates or predictions based on in vitro physiological measurements. These measurements have been shown to significantly overestimate the expected maximum speed of what were thought to be the fastest swimmers, like billfish. 
First, these fish have lengths $L>L_c$ and consequently their tail beat frequencies are significantly smaller than the maximum frequency $f_0$ expected from the muscles: we predict that a 2 m-long swimmer with $f_0=20\ \mathrm{ Hz}$ would swim with a maximum tail beat frequency four times smaller, below 5 Hz (Fig. \ref{fig:FrequencyVsLength}), and thus would have a maximum speed reduced by a factor of 4 in comparison to predictions based on the twitch contraction time only. 
Second, recent estimates based on measurements of twitch contraction times of anaerobic muscles also provide upper bounds lower than $10\ \mathrm{m.s^{-1}}$ for billfish and other large marine predatory fish \cite{svendsen_maximum_2016}. Actually, reported values of maximum speeds agree with this argument and refute some incredibly huge values that had been estimated for animals of this size. For marlins, direct measurements using speedometers showed that the observed maximum swimming speed was around $2.25\ \mathrm{m.s^{-1}}$ \cite{block1992direct}, considerably lower than the estimates of $30\ \mathrm{m.s^{-1}}$\cite{lane1941fast}. Burst speeds of sailfish also show values around 8 m.s$^{-1}$ measured with high-speed video and accelerometry \cite{marras2015not}, a value much smaller than 30 m.s$^{-1}$ \cite{lane1941fast}. In addition, there is some theoretical evidence that the maximum speed should be smaller than 15 m.s$^{-1}$, because cavitation should appear at greater speeds, which should damage the flesh of the swimmer \cite{iosilevskii2008speed}. In Fig. \ref{fig:VelocityVsLength}, we have used open translucent squares to represent data that were not actual measurements but estimates, or that were taken from non-peer-reviewed studies (Methods \ref{RefHighVelocity}). If we remove these points from the analysis, we find that the two upper bounds of the data sets match very well, with the exception of four data points obtained for tuna and barracuda that are still significantly faster than our fit of the fast bound (open opaque squares in Fig. \ref{fig:VelocityVsLength}). 
Note that all of these points were measured using rod-mounted devices that measure the speed at which the line is pulled from the reel when a fish is hooked and pulling on the line (\cite{bainbridge1958speed,walters1964measurements, bonner2015size} and Methods \ref{RefHighVelocity}). Given the large fluctuations in the measurements made with this method \cite{walters1964measurements}, it is likely that it overestimates the maximum speed, which is also supported by the fact that barracuda and tuna do not show particularly high maximum frequencies \cite{svendsen_maximum_2016}.

\item
For $L \gtrsim 5$ m, we attribute the discrepancy to a lack of tail beat frequency measurements for very long swimmers, typically cetaceans, at a burst level of activity. This would explain the jump in frequencies around $L \sim 5$ m  in Fig. \ref{fig:FrequencyVsLength} for the fast bound. In this figure, the measured frequencies above $L = 5$ m correspond to swimming speeds between 1 and 4 m.s$^{-1}$ \cite{sato2007stroke, gough2019scaling, gough2021scaling}. Swimming speeds up to 10 m.s$^{-1}$ were recorded in sperm whales \cite{aoki2007measurement}, for which tail beat frequencies were unfortunately not measured, but which should be consistently higher than those plotted in Fig. \ref{fig:FrequencyVsLength} for the same length. Unlike sperm whales, killer whales and some other large marine predators, most cetaceans are filter feeders and do not have predators due to their size. Therefore, they do not often need to move at maximum speeds, which would favor data closer to the slow bound than the fast bound.
\end{enumerate}
Following these considerations, it is reasonable to consider that the maximum speed is constant for $L \gtrsim 0.5 - 1$ m, with a typical value around 5--10 m.s$^{-1}$, in very good agreement with the prediction $0.7 f_0 L_c$ inferred from the fit of the fast bound. From the definition of $L_c$, it means that $U \sim \sqrt{\sigma_0/\rho}$ for $L > L_c$ and that swimming speed is directly associated with the maximum stress generated by the muscles to push the surrounding water. In their model, Hirt et al. state that heavier (and consequently longer) animals need more time to accelerate to achieve maximum speed and this fact would prevent the heaviest animals from being the fastest. Here we suggest that the effect of a finite acceleration time would be a second-order effect, unlike the other locomotion modes running and flying \cite{hirt2017general}.

\section{Discussion}

From the results, we conclude that length and activity level determine swimming frequency and speed at the leading order. The data collected on natural swimmers are explained using a simple model that accounts for biological characteristics, through Hill's muscle model, as well as the interaction of the undulating swimmer with its environment. This model requires only a few parameters that could be refined in the future to account for specific characteristics of each swimmer (body temperature, swimming gait, etc.). This work broadens our understanding of animal locomotion but should also help in designing biomimetic and autonomous swimming robots \cite{lebastard_reactive_2016,zhu_tuna_2019,sanchez-rodriguez_proprioceptive_2021,thandiackal_emergence_2021,lee_autonomously_2022}, with the constraint that artificial swimmers have their own internal characteristics that replace the  biological ones discussed in the present study.

Our study highlights a crossover at a length $L_c \sim 0.5 - 1$ m that  separates two limits: while small swimmers are constrained by biology only, large swimmers are constrained by their environment as well. For a given activity level, different scaling laws are found for swimming frequency and speed in the two limits. This should also be the case for other quantities, such as the muscle power for locomotion. For this quantity, our model predicts scaling laws in $L^{5}$ and $L^{2}$ for very small and very large swimmers, respectively (Methods \ref{Methods:Hill}). Measurements of oxygen consumption \cite{brett1964respiratory,brett_metabolic_1972} over a wide range of lengths and activity levels might be a way to test these predictions. In the framework of the model, $L_c$ also marks a significant change in the way muscles are used. Small swimmers use muscles at their maximum speed but negligible force in comparison to their maximal capabilities (Methods \ref{Methods:Hill}). Large swimmers exhibit the opposite behavior. Given that muscle power for locomotion scales as the product of force, frequency and length, we expect  muscle power for locomotion to be negligible in these two limits in comparison to the maximum power available. Remarkably, only intermediate fish, with lengths around $L_c$, would use the full capacity of muscle power to undulate and move through water. In light of this comment, we can question in the future whether intermediate fish are more likely to use economic locomotion strategies (e.g., intermittent swimming \cite{li2021burst}, schooling \cite{marras_fish_2015}, etc.) compared to very small and large swimmers.

\section{Methods}
\label{Methods}

\subsection{Data and allometry plots}
\label{Methods:Data}

\begin{figure}[H]
a) \centering\includegraphics[width=0.45\textwidth]{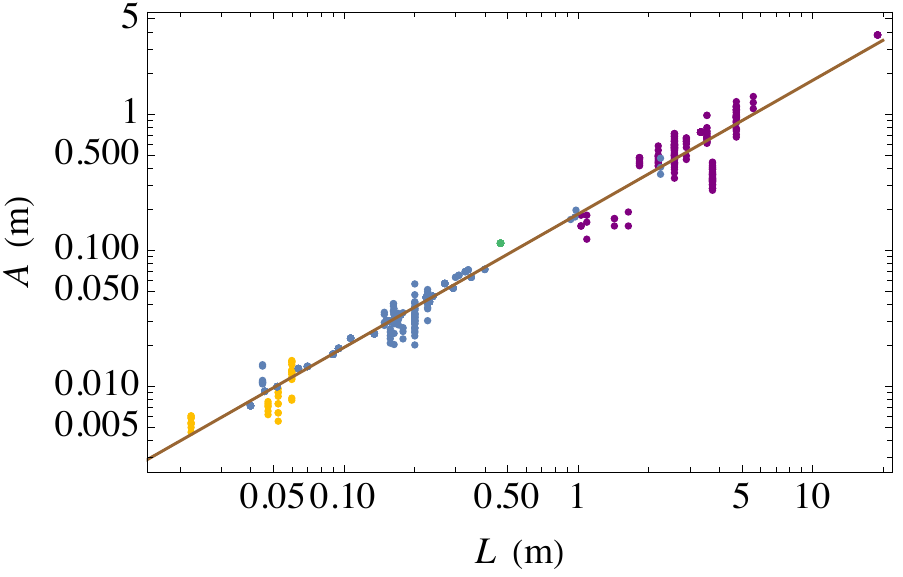}
b) 
\centering\includegraphics[width=0.45\textwidth]{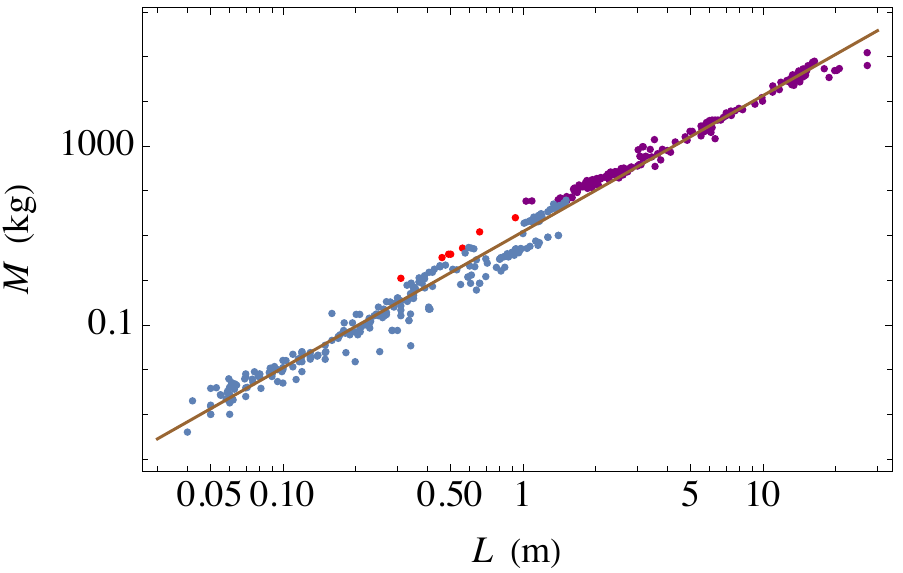}
\caption{a) Tail beat amplitude and b) animal mass as a function of animal length. The solid lines represent the best power-law fits of the data.} 
\label{fig:AmplitudeVsLength}
\end{figure}

We retrieved the length-frequency data available in the literature for a total of 1202 animals, with a range of different species, morphologies and sizes. We  regrouped the data according to the classical division of vertebrates: amphibians, fish, birds, reptiles  and mammals. In figures, these data points are displayed in yellow, blue, red, green and purple, respectively. 
In the cases where the length data were not reported \cite{sato2007stroke} but instead the mass of the animals, the length was deduced through the allometric relation deduced in Fig. \ref{fig:AmplitudeVsLength}b $L = 0.44 M^{0.33}$ ($L$ in meters and $M$ in kilograms), assuming geometric similarity. Additionally, we recovered length-amplitude and length-mass data. The references we used are provided in Tab. \ref{table:references}.

\begin{table}[H]
\centering
\begin{tabular}{c|ccc|}
\cline{2-4}
  & \multicolumn{3}{c|}{References}                                        \\ \cline{2-4} 
     & \multicolumn{1}{c|}{Frequency vs length or mass} & \multicolumn{1}{c|}{Amplitude vs length or mass} & Mass vs length \\ \hline
\multicolumn{1}{|c|}{Amphibians}   & \multicolumn{1}{c|}{\cite{wassersug1985kinematics}}          & \multicolumn{1}{c|}{\cite{wassersug1985kinematics}}          &   -   \\ \hline
\multicolumn{1}{|c|}{Fish}       & \multicolumn{1}{c|}{\cite{bainbridge1958speed,hunter1971swimming,webb_kinematics_1986,webb_steady_nodate,videler_fast_1983,rosenberger_functional_nodate,magnuson1966courtship,mueller2010tail,thiem2015accelerometer,webb1982swimming,graham1990aspects,marras2015not,watanabe_slowest_2012}}          & \multicolumn{1}{c|}{\cite{bainbridge1958speed,hunter1971swimming,webb_kinematics_1986,webb_steady_nodate,videler_fast_1983,rosenberger_functional_nodate,graham1990aspects}}          &   \cite{bainbridge1958speed,yan2013interspecific,malik2020ontogeny,dissanayake2008fishery,mehanna2021length,motta2013technical}    \\ \hline
\multicolumn{1}{|c|}{Reptiles} & \multicolumn{1}{c|}{\cite{fish1984kinematics}}          & \multicolumn{1}{c|}{\cite{fish1984kinematics}}          &   -   \\ \hline
\multicolumn{1}{|c|}{Birds}      & \multicolumn{1}{c|}{\cite{sato2010scaling,clark1979kinematics}}          & \multicolumn{1}{c|}{-}          &  \cite{sato2010scaling}    \\ \hline
\multicolumn{1}{|c|}{Mammals}    & \multicolumn{1}{c|}{\cite{fish_comparative_nodate,kojeszewski2007swimming,sato2007stroke,goldbogen2006kinematics,rohr2004strouhal,gough2021scaling,fish_kinematics_nodate}}          & \multicolumn{1}{c|}{\cite{fish_comparative_nodate,kojeszewski2007swimming,rohr2004strouhal,fish_kinematics_nodate}}          &  \cite{nishiwaki1950body,kojeszewski2007swimming,robeck2005reproduction,christiansen2019estimating,fish_comparative_nodate,fish_kinematics_nodate}    \\ \hline
\end{tabular}
\caption{References reporting relations between length, frequency,  amplitude or mass of swimmers.
}
\label{table:references}
\end{table}

Our approach is based on two relationships verified through the data in Tab. \ref{table:references}. 
First, we plotted $A$ versus $L$ in Fig. \ref{fig:AmplitudeVsLength}a and verified that $A$ is proportional to $L$, as originally shown by Bainbridge \cite{bainbridge1958speed}, but extended here to 359 different specimens and four orders of magnitude in length.
In fact the best fit with a power law gives $A = (0.183\pm0.002) L^{0.981\pm 0.005}$, which shows that both quantities are proportional. The best proportionality relation gives  $A = (0.187\pm0.001) L$. 
The geometric similarity for aquatic animals is shown in Fig. \ref{fig:AmplitudeVsLength}b for 432 different individuals. The best fit of the data gives  $M = \left( 12.14 \pm 0.33 \right) L^{3.04 \pm 0.02}$, with $M$ in kilograms and $L$ in meters, which is consistent with the geometric similarity characterized by an exponent of 3. This law extends Economos's relation ($M =11.27 L^{ 2.95}$ \cite{economos1983elastic}) over four orders of magnitude in length or ten in mass. Forcing the exponent of the relation to be exactly 3 gives  $M = (12.09\pm0.33) L^3$ or its dimensional homogeneous form $M = (0.0121 \pm 0.0003) \rho L^3$ with $\rho = 1000$ kg.m$^{-3}$.

\subsection{Hill's muscle model}
\label{Methods:Hill}

\begin{figure}[H]
\centering\includegraphics[width=0.95\textwidth]{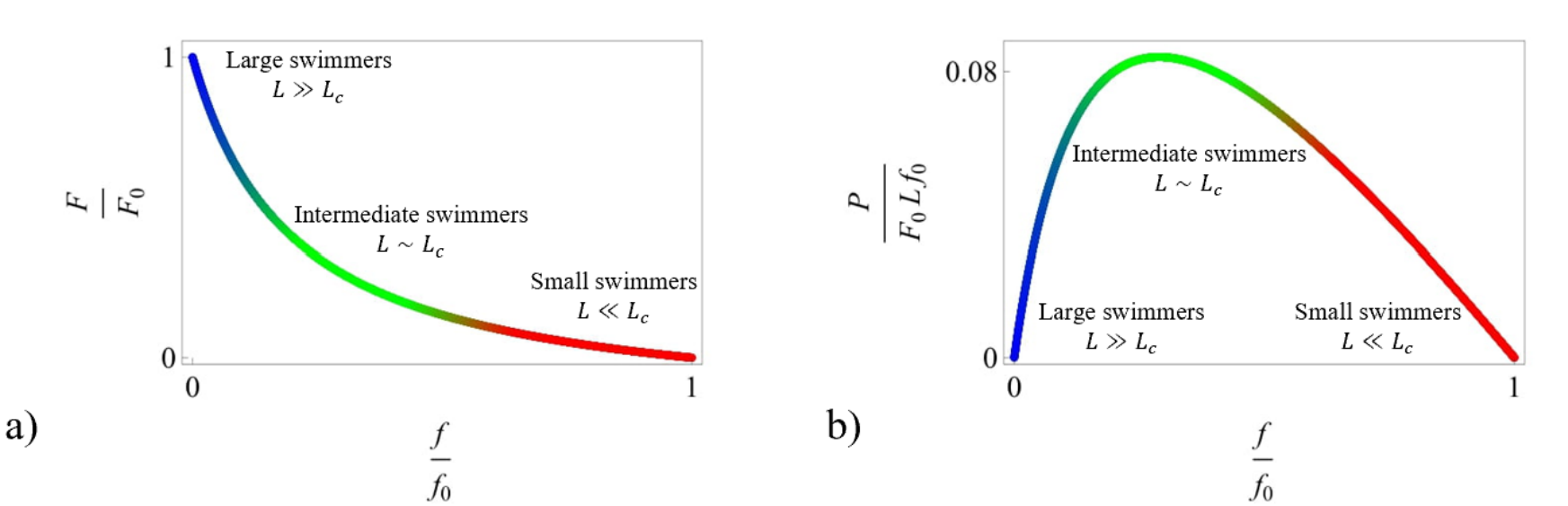}
\caption{a) Muscle force and b) muscle power for locomotion as a function of tail beat frequency, as predicted by Hill's muscle model. The curves are drawn with $\kappa = 5$ and the quantities are represented in their dimensionless form.
} 
\label{fig:ForcePowerHill}
\end{figure}

Hill's equations for tetanized muscle contraction (Eq. (\ref{HillEquation})) can be rewritten to express the force in the muscle $F$ as a function of the swimming frequency $f$: 
\begin{equation}
    \frac{F}{F_0} = \frac{1 - \frac{f}{f_0} }{1 + \kappa \frac{f}{f_0} } .
\end{equation}
$F_0$ is the maximum isometric force generated in the muscle and $f_0$ is the maximum tail beat frequency. In its dimensionless form, the force $F/F_0$ is a decreasing and convex function of the frequency $f/f_0$, whose degree of curvature is quantified by the parameter $\kappa$ (Fig. \ref{fig:ForcePowerHill}a). From our analysis, very small animals ($L \ll L_c$) swim at maximum frequency and negligible force ($f = f_0$ and $F \ll F_0$), while very large animals ($L \gg L_c$) swim at negligible frequency and maximum force ($f\ll f_0$ and $F = F_0$). Intermediate sized animals ($L \sim L_c$) swim  at intermediate frequency and force ($f \lessapprox f_0$ and $F \lessapprox F_0$).

An estimate of muscle power for locomotion $P$ is obtained by multiplying the muscle force by $L f$, an estimate of the speed of muscle contraction. $P \approx F L f$ can be expressed as a function of the tail beat frequency following Hill's equation:
\begin{equation}
    \frac{P}{F_0 L f_0} = \frac{1 - \frac{f}{f_0} }{1 + \kappa \frac{f}{f_0} }   \frac{f}{f_0} .
\end{equation}
The power is drawn in Fig. \ref{fig:ForcePowerHill}b and $F_0 L f_0$ gives an estimate of the maximum power that can be delivered by the muscle. This graph highlights the fact that very small and large swimmers do not use the full capacity of muscle power, unlike intermediate sized swimmers ($L \sim L_c$).

It is also possible to derive the expected scaling laws for muscle power as a function of length. To do this, we take $F \sim \rho L^4 f^2$ from the interaction of the undulating swimmer with its environment and $f=f_0$ or $\sqrt{\sigma_0/\rho}/L$
for very small or large swimmers, respectively. This gives the following scaling laws in these two limits:
\begin{eqnarray}
    \label{eq:asymptiticsPa}
    P &\sim& \rho f_0^3 L^5, \quad  \mathrm{if}\quad L \ll L_c\\
    \label{eq:asymptiticsPb}
    P &\sim& \rho^{-1/2}  \sigma_0^{3/2} L^2, \quad  \mathrm{if}\quad L \gg L_c.
\end{eqnarray}

\subsection{Characterization of the burst and sustained activity levels}
\label{Methods:FitActivityRegimes}

All measurements are located within a band in the $L - f$ plane. The fast and slow bounds corresponding to the burst and sustained activity levels, respectively, are defined as follows. First, the length axis is divided into $N$ intervals equally distributed in logarithmic scale. Second, in each interval, the maximum and minimum frequencies are identified, as well as  all data points that are found between 90\% and 100\% of the maximum frequency value and  between 100\% and 110\% of the minimum frequency value. Finally, these two sets of points are averaged independently to give one point for each bound inside each interval. 

For each bound, the three parameters of the model ($L_c$, $f_0$ and $\kappa$) are obtained by fitting the $N$ averaged points (one per interval) with Eq. (\ref{eqBalance}) using the least absolute deviations (LAD) method. The robustness and precision of the fit is probed by varying $N$ and taking the mean and standard deviation to characterize each parameter. 
The number of intervals, $N$, was varied between 10 and 50. This range ensured a statistically significant number of points while at the same time avoiding empty intervals. In practice, we selected  nine values uniformly distributed between these two extreme values.

\subsection{Filtering of maximum speed data}
\label{RefHighVelocity}

We investigated the origin of the data gathered by Hirt et al. \cite{hirt2017general} to establish objective criteria on data selection. Of all the measures reported in the study, we found three classes of data that were not as reliable as the others. 
First, we identified data coming from non-peer reviewed papers. 
Second, some of the data are only estimates, not actual measurements. 
Third, we found that all data obtained with rod-mounted devices are above the main trend, which could be artificial due to the high fluctuations in these measurements \cite{walters1964measurements}. 
All these three classes of data points are summarized in Tab. \ref{table:maxVelocity}.

\begin{table}[H]
\centering
\begin{tabular}{|c|c|c|c|c|c|}
\hline
Animal & Mass (kg) & Length (m) &  \shortstack{ Maximum \\ speed (m.s$^{-1}$)} & \shortstack{ Speed \\ reference} & \shortstack{ Dismission \\ reason}  \\ \hline
Aptenodytes patagonicus &  14   &  1.05      &       3.36   &     \cite{webPageSpeedAnimals}  & Non-peer reviewed     \\ \hline
Pygoscelis antarcticus &  4.5   &   0.72     &       8.89   &     \cite{webPageSpeedAnimals}   & Non-peer reviewed    \\ \hline
Pygoscelis papua &  8.2   &  0.88      &       10   &     \cite{webPageSpeedAnimals}   & Non-peer reviewed    \\ \hline
  Acanthocybium solandri &  13.31   & 1.03       &       21.4   &     \cite{bonner2015size}    & Rod and reel   \\ \hline
  Acanthocybium solandri  &  16.64  &  1.11      &     21.39     &   \cite{webPageSpeedAnimals}    & Non-peer reviewed     \\ \hline
  Carcharodon carcharias  &  800  &  4.04      &     11.11     &   \cite{webPageSpeedAnimals}   & Non-peer reviewed     \\ \hline
  Galeocerdo cuvier  &  550  &   3.57     &     8.89     &   \cite{webPageSpeedAnimals}   & Non-peer reviewed     \\ \hline
  Istiompax indica  &  150  &  2.32      &     36.11     &   \cite{block1992direct}    & Estimation    \\ \hline
  Istiophorus albicans  &  90  &  1.95      &     30.56     &   \cite{webPageSpeedAnimals}  & Non-peer reviewed      \\ \hline
  Istiophorus albicans  &  90  &   1.95     &     30     &   \cite{block1992direct}    & Estimation    \\ \hline
  Isurus oxyrinchus  &  105  &   2.06     &     18.8     &   \cite{wikipediaMakoShark} & Non-peer reviewed       \\ \hline
  Isurus oxyrinchus  &  300  &    2.92    &     13.89     &   \cite{webPageSpeedAnimals}  & Non-peer reviewed      \\ \hline
  Makaira nigricans  &  153.5  &  2.33      &     20.83     &   \cite{block1992direct}    & Estimation  \\ \hline
  Sphyranea argentea  &  4.5  &   0.72     &     12.22     &   \cite{webPageSpeedAnimals}   & Non-peer reviewed    \\ \hline
  Sphyranea barracuda  &  26.56  &   1.29     &     12.19     &   \cite{bainbridge1958speed}  & Rod and reel     \\ \hline
  Tetrapturus audax  &  163  &    2.38    &     22.5     &   \cite{webPageSpeedAnimals}  & Non-peer reviewed     \\ \hline
  Thunnus albacares  &  13.11  &   1.03     &     20.83     &   \cite{walters1964measurements} & Rod and reel     \\ \hline
  Thunnus orientalis  &  250  &   2.74     &     19.44     &   \cite{webPageSpeedAnimals}   & Non-peer reviewed   \\ \hline
  Thunnus thynnus  &  27.22  &   1.31     &     19.67     &   \cite{bainbridge1958speed}    & Rod and reel  \\ \hline
  Xiphias gladius  &  98  &   2.01     &     26.94     &   \cite{webPageSpeedAnimals}  & Non-peer reviewed    \\ \hline
   Balaenoptera  musculus &  108400  &   20.75     &     10.30     &   \cite{hill_dimensions_1950} & Estimation     \\ \hline
    Delphinus delphis &  95.32  &   1.99     &     10.30     &   \cite{hill_dimensions_1950} & Estimation      \\ \hline
  Enhydra lutris  &  30  &   1.35     &     2.5     &   \cite{webPageSpeedAnimals}  & Non-peer reviewed    \\ \hline
  Megaptera novaeanglia  &  30000  &  13.53      &     7.5     &   \cite{webPageSpeedAnimals}  & Non-peer reviewed    \\ \hline
  Orcinus orca  &  4100  &  6.97      &     13.33     &   \cite{webPageSpeedAnimals}   & Non-peer reviewed   \\ \hline
  Orcinus orca  &  4300  &   7.09     &     15.4     &   \cite{webPageOrcinusOrca}    & Non-peer reviewed  \\ \hline
  Pusa hispida  &  88.07  &  1.94      &     8.33     &   \cite{webPageRingedSeal}    & Non-peer reviewed  \\ \hline
  Tursiops truncatus  &  250  &  2.74      &     9.72     &   \cite{webPageSpeedAnimals}  & Non-peer reviewed   \\ \hline
  Zalophus californianus  &  158  &  2.36      &     11.11     &   \cite{webPageSeaLion}   & Non-peer reviewed   \\ \hline
  Dermochelys coriacea  &  420  &   3.26     &     9.8     &   \cite{webPageOrcinusOrca}  & Non-peer reviewed    \\ \hline
\end{tabular}
\caption{Data used in Hirt et al. \cite{hirt2017general} that are either provided by non-peer-reviewed articles, given as estimates, or obtained using rod-mounted devices. We list the type of animal together with the corresponding mass and speed, the length using Methods \ref{Methods:Data} if not provided, and references.
}
\label{table:maxVelocity}
\end{table}

\paragraph{Acknowledgements} 
We are grateful to Fran\c cois Gallaire, Guillaume Allibert and José Luis Trejo for enlightning discussions. This work was supported by the French government, through the UCAJEDI Investments in the Future project of the National Research Agency (ANR-15-IDEX-01).


\bibliographystyle{unsrt}
\bibliography{./referencesAllometry}

\begin{thebibliography}{10}

\bibitem{lauder2005hydrodynamics}
George~V Lauder and Eric~D Tytell.
\newblock Hydrodynamics of undulatory propulsion.
\newblock {\em Fish physiology}, 23:425--468, 2005.

\bibitem{sfakiotakis1999review}
Michael Sfakiotakis, David~M Lane, and J~Bruce~C Davies.
\newblock Review of fish swimming modes for aquatic locomotion.
\newblock {\em IEEE Journal of oceanic engineering}, 24(2):237--252, 1999.

\bibitem{videler_fish_1993}
J.~J. Videler.
\newblock {\em Fish {Swimming}}.
\newblock Springer Netherlands, 1993.

\bibitem{DiSanto2021}
Valentina~Di Santo, Elsa Goerig, Dylan~K. Wainwright, Otar Akanyeti, James~C.
  Liao, Theodore Castro-Santos, and George~V. Lauder.
\newblock Convergence of undulatory swimming kinematics across a diversity of
  fishes.
\newblock {\em Proceedings of the National Academy of Sciences},
  118(49):e2113206118, 2021.

\bibitem{bainbridge1958speed}
Richard Bainbridge.
\newblock The speed of swimming of fish as related to size and to the frequency
  and amplitude of the tail beat.
\newblock {\em Journal of experimental biology}, 35(1):109--133, 1958.

\bibitem{rohr2004strouhal}
Jim~J Rohr and Frank~E Fish.
\newblock Strouhal numbers and optimization of swimming by odontocete
  cetaceans.
\newblock {\em Journal of Experimental Biology}, 207(10):1633--1642, 2004.

\bibitem{hunter1971swimming}
JR~Hunter.
\newblock Swimming speed, tail beat frequency, tail beat amplitude and size in
  jack mackerel, trachurus symmetricus, and other fishes.
\newblock {\em Fish. Bull.}, 69:253--266, 1971.

\bibitem{saadat2017rules}
M~Saadat, Frank~E Fish, AG~Domel, V~Di~Santo, GV~Lauder, and H~Haj-Hariri.
\newblock On the rules for aquatic locomotion.
\newblock {\em Physical Review Fluids}, 2(8):083102, 2017.

\bibitem{triantafyllou1993optimal}
George~S Triantafyllou, Michael~S Triantafyllou, and Mark~A Grosenbaugh.
\newblock Optimal thrust development in oscillating foils with application to
  fish propulsion.
\newblock {\em Journal of Fluids and Structures}, 7(2):205--224, 1993.

\bibitem{gazzola2014scaling}
Mattia Gazzola, M{\'e}d{\'e}ric Argentina, and Lakshminarayanan Mahadevan.
\newblock Scaling macroscopic aquatic locomotion.
\newblock {\em Nature Physics}, 10(10):758--761, 2014.

\bibitem{videler_fish_1991}
J.~J. Videler and C.~S. Wardle.
\newblock Fish swimming stride by stride: speed limits and endurance.
\newblock {\em Rev Fish Biol Fisheries}, 1(1):23--40, September 1991.

\bibitem{videler_differences_1985}
J.~Videler and P.~Kamermans.
\newblock Differences between upstroke and downstroke in swimming dolphins.
\newblock {\em Journal of Experimental Biology}, 119(1):265--274, November
  1985.

\bibitem{curren_designs_1992}
Kristina~Charlotte Curren.
\newblock Designs for swimming : morphometrics and swimming dynamics of several
  cetacean species.
\newblock Master's thesis, Memorial University of Newfoundland, 1992.

\bibitem{svendsen_maximum_2016}
Morten B.~S. Svendsen, Paolo Domenici, Stefano Marras, Jens Krause, Kevin~M.
  Boswell, Ivan Rodriguez-Pinto, Alexander D.~M. Wilson, Ralf H. J.~M. Kurvers,
  Paul~E. Viblanc, Jean~S. Finger, and John~F. Steffensen.
\newblock Maximum swimming speeds of sailfish and three other large marine
  predatory fish species based on muscle contraction time and stride length: a
  myth revisited.
\newblock {\em Biology Open}, 5(10):1415--1419, August 2016.

\bibitem{bejan2006unifying}
Adrian Bejan and James~H Marden.
\newblock Unifying constructal theory for scale effects in running, swimming
  and flying.
\newblock {\em Journal of Experimental Biology}, 209(2):238--248, 2006.

\bibitem{sato2007stroke}
Katsufumi Sato, Yutaka Watanuki, Akinori Takahashi, Patrick~JO Miller, Hideji
  Tanaka, Ryo Kawabe, Paul~J Ponganis, Yves Handrich, Tomonari Akamatsu, Yuuki
  Watanabe, et~al.
\newblock Stroke frequency, but not swimming speed, is related to body size in
  free-ranging seabirds, pinnipeds and cetaceans.
\newblock {\em Proceedings of the Royal Society B: Biological Sciences},
  274(1609):471--477, 2007.

\bibitem{watanabe_slowest_2012}
Yuuki~Y. Watanabe, Christian Lydersen, Aaron~T. Fisk, and Kit~M. Kovacs.
\newblock The slowest fish: {Swim} speed and tail-beat frequency of {Greenland}
  sharks.
\newblock {\em Journal of Experimental Marine Biology and Ecology},
  426-427:5--11, September 2012.

\bibitem{bale2014energy}
Rahul Bale, Max Hao, Amneet Pal~Singh Bhalla, and Neelesh~A Patankar.
\newblock Energy efficiency and allometry of movement of swimming and flying
  animals.
\newblock {\em Proceedings of the National Academy of Sciences},
  111(21):7517--7521, 2014.

\bibitem{gough2019scaling}
William~T Gough, Paolo~S Segre, KC~Bierlich, David~E Cade, Jean Potvin, Frank~E
  Fish, Julian Dale, Jacopo di~Clemente, Ari~S Friedlaender, David~W Johnston,
  et~al.
\newblock Scaling of swimming performance in baleen whales.
\newblock {\em Journal of Experimental Biology}, 222(20):jeb204172, 2019.

\bibitem{zhao2012effects}
Wen-Wen Zhao, Xu~Pang, Jiang-Lan Peng, Zhen-Dong Cao, and Shi-Jian Fu.
\newblock The effects of hypoxia acclimation, exercise training and fasting on
  swimming performance in juvenile qingbo (spinibarbus sinensis).
\newblock {\em Fish physiology and biochemistry}, 38(5):1367--1377, 2012.

\bibitem{allen2006effects}
Peter~J Allen, Brian Hodge, Inge Werner, and Joseph~J Cech, Jr.
\newblock Effects of ontogeny, season, and temperature on the swimming
  performance of juvenile green sturgeon (acipenser medirostris).
\newblock {\em Canadian Journal of Fisheries and Aquatic Sciences},
  63(6):1360--1369, 2006.

\bibitem{makiguchi2017sex}
Yuya Makiguchi, Hisaya Nii, Katsuya Nakao, and Hiroshi Ueda.
\newblock Sex differences in metabolic rate and swimming performance in pink
  salmon (o ncorhynchus gorbuscha): the effect of male secondary sexual traits.
\newblock {\em Ecology of Freshwater Fish}, 26(2):322--332, 2017.

\bibitem{wu1977introduction}
Theodore~Y Wu.
\newblock Introduction to the scaling of aquatic animal locomotion.
\newblock Technical report, CALIFORNIA INST OF TECH PASADENA, 1977.

\bibitem{brett1964respiratory}
J~Roland Brett.
\newblock The respiratory metabolism and swimming performance of young sockeye
  salmon.
\newblock {\em Journal of the Fisheries Board of Canada}, 21(5):1183--1226,
  1964.

\bibitem{brett_metabolic_1972}
J.~R. Brett.
\newblock The metabolic demand for oxygen in fish, particularly salmonids, and
  a comparison with other vertebrates.
\newblock {\em Respiration Physiology}, 14(1):151--170, 1972.

\bibitem{videler_fast_1983}
JJ~Videler and F~Hess.
\newblock Fast continuous swimming of two pelagic predators, saithe
  ({P}ollachius virens) and mackerel ({S}comber scombrus): a kinematic
  analysis.
\newblock {\em Journal of experimental biology}, 109(1):209--228, 1984.

\bibitem{fish_comparative_nodate}
Frank~E Fish.
\newblock Comparative kinematics and hydrodynamics of odontocete cetaceans:
  morphological and ecological correlates with swimming performance.
\newblock {\em The Journal of experimental biology}, 201(20):2867--2877, 1998.

\bibitem{marras2015not}
Stefano Marras, Takuji Noda, John~F Steffensen, Morten~BS Svendsen, Jens
  Krause, Alexander~DM Wilson, Ralf~HJM Kurvers, James Herbert-Read, Kevin~M
  Boswell, and Paolo Domenici.
\newblock Not so fast: swimming behavior of sailfish during predator--prey
  interactions using high-speed video and accelerometry.
\newblock {\em Integrative and comparative biology}, 55(4):719--727, 2015.

\bibitem{shadwick2006fish}
Robert~E Shadwick and George~V Lauder.
\newblock {\em Fish physiology: fish biomechanics}.
\newblock Elsevier, 2006.

\bibitem{hill1938heat}
Archibald~Vivian Hill.
\newblock The heat of shortening and the dynamic constants of muscle.
\newblock {\em Proceedings of the Royal Society of London. Series B-Biological
  Sciences}, 126(843):136--195, 1938.

\bibitem{hill_dimensions_1950}
A.~V. HILL.
\newblock {THE} {DIMENSIONS} {OF} {ANIMALS} {AND} {THEIR} {MUSCULAR}
  {DYNAMICS}.
\newblock {\em Science Progress (1933- )}, 38(150):209--230, 1950.

\bibitem{wardle_limit_1975}
C.~S. Wardle.
\newblock Limit of fish swimming speed.
\newblock {\em Nature}, 255(5511):725--727, June 1975.

\bibitem{matsuura_muscle_2020}
Yuiko Matsuura, Naoto Matsunaga, Satoshi Iizuka, Hiroshi Akuzawa, and Koji
  Kaneoka.
\newblock Muscle {Synergy} of the {Underwater} {Undulatory} {Swimming} in
  {Elite} {Male} {Swimmers}.
\newblock {\em Frontiers in Sports and Active Living}, 2:62, June 2020.

\bibitem{landau1986Elasticity}
Lev~Davidovich Landau, Evgenij~M Lif{\v{s}}ic, Evegnii~Mikhailovich Lifshitz,
  Arnold~Markovich Kosevich, and Lev~Petrovich Pitaevskii.
\newblock {\em Theory of elasticity: volume 7}, volume~7.
\newblock Elsevier, 1986.

\bibitem{gazzola2015gait}
Mattia Gazzola, M{\'e}d{\'e}ric Argentina, and Lakshminarayanan Mahadevan.
\newblock Gait and speed selection in slender inertial swimmers.
\newblock {\em Proceedings of the National Academy of Sciences},
  112(13):3874--3879, 2015.

\bibitem{paraz2016thrust}
Florine Paraz, Lionel Schouveiler, and Christophe Eloy.
\newblock Thrust generation by a heaving flexible foil: Resonance,
  nonlinearities, and optimality.
\newblock {\em Physics of Fluids}, 28(1):011903, 2016.

\bibitem{hoover2018swimming}
Alexander~P Hoover, Ricardo Cortez, Eric~D Tytell, and Lisa~J Fauci.
\newblock Swimming performance, resonance and shape evolution in heaving
  flexible panels.
\newblock {\em Journal of Fluid Mechanics}, 847:386--416, 2018.

\bibitem{Fitts1991111}
Robert~H. Fitts, Kerry~S. McDonald, and Jane~M. Schluter.
\newblock The determinants of skeletal muscle force and power: Their
  adaptability with changes in activity pattern.
\newblock {\em Journal of Biomechanics}, 24(SUPPL. 1):111--122, 1991.
\newblock Cited by: 120.

\bibitem{rome1988animals}
Lawrence~C Rome, Roel~P Funke, R~McNeill Alexander, Gordon Lutz, Hugh Aldridge,
  Frank Scott, and Marvin Freadman.
\newblock Why animals have different muscle fibre types.
\newblock {\em Nature}, 335(6193):824--827, 1988.

\bibitem{johnston_thermal_1984}
Ian~A. Johnston and Richard Brill.
\newblock Thermal dependence of contractile properties of single skinned muscle
  fibres from {Antarctic} and various warm water marine fishes including
  {Skipjack} {Tuna} ({Katsuwonus} pelamis) and {Kawakawa} ({Euthynnus}
  affinis).
\newblock {\em J Comp Physiol B}, 155(1):63--70, January 1984.

\bibitem{altringham_pca-tension_1982}
J.~D. Altringham and I.~A. Johnston.
\newblock The {pCa}-tension and force-velocity characteristics of skinned
  fibres isolated from fish fast and slow muscles.
\newblock {\em The Journal of Physiology}, 333(1):421--449, 1982.

\bibitem{johnston1985force}
IA~Johnston, Bruce~D Sidell, and WR~Driedzic.
\newblock Force-velocity characteristics and metabolism of carp muscle fibres
  following temperature acclimation.
\newblock {\em Journal of Experimental Biology}, 119(1):239--249, 1985.

\bibitem{wilkie1949relation}
DR~Wilkie.
\newblock The relation between force and velocity in human muscle.
\newblock {\em The Journal of physiology}, 110(3-4):249, 1949.

\bibitem{johnston1985sustained}
IA~Johnston.
\newblock Sustained force development: specializations and variation among the
  vertebrates.
\newblock {\em Journal of Experimental Biology}, 115(1):239--251, 1985.

\bibitem{hooper2005invertebrate}
Scott~L Hooper and Jeffrey~B Thuma.
\newblock Invertebrate muscles: muscle specific genes and proteins.
\newblock {\em Physiological reviews}, 2005.

\bibitem{wardle_effects_1980}
C.~S. Wardle.
\newblock Effects of {Temperature} on the {Maximum} {Swimming} {Speed} of
  {Fishes}.
\newblock In M.~A. Ali, editor, {\em Environmental {Physiology} of {Fishes}},
  {NATO} {Advanced} {Study} {Institutes} {Series}, pages 519--531. Springer US,
  Boston, MA, 1980.

\bibitem{johnson_thermal_1995}
T~Johnson and A~Bennett.
\newblock The thermal acclimation of burst escape performance in fish: an
  integrated study of molecular and cellular physiology and organismal
  performance.
\newblock {\em Journal of Experimental Biology}, 198(10):2165--2175, October
  1995.

\bibitem{sato2010scaling}
Katsufumi Sato, Kozue Shiomi, Yuuki Watanabe, Yutaka Watanuki, Akinori
  Takahashi, and Paul~J Ponganis.
\newblock Scaling of swim speed and stroke frequency in geometrically similar
  penguins: they swim optimally to minimize cost of transport.
\newblock {\em Proceedings of the Royal Society B: Biological Sciences},
  277(1682):707--714, 2010.

\bibitem{hirt2017general}
Myriam~R Hirt, Walter Jetz, Bj{\"o}rn~C Rall, and Ulrich Brose.
\newblock A general scaling law reveals why the largest animals are not the
  fastest.
\newblock {\em Nature Ecology \& Evolution}, 1(8):1116--1122, 2017.

\bibitem{block1992direct}
Barbara~A Block, David Booth, and Francis~G Carey.
\newblock Direct measurement of swimming speeds and depth of blue marlin.
\newblock {\em Journal of Experimental Biology}, 166(1):267--284, 1992.

\bibitem{lane1941fast}
FW~Lane.
\newblock How fast do fish swim.
\newblock {\em Country Life (London)}, 90:534--535, 1941.

\bibitem{iosilevskii2008speed}
G~Iosilevskii and D~Weihs.
\newblock Speed limits on swimming of fishes and cetaceans.
\newblock {\em Journal of The Royal Society Interface}, 5(20):329--338, 2008.

\bibitem{walters1964measurements}
Vladimir Walters and Harry~L Fierstine.
\newblock Measurements of swimming speeds of yellowfin tuna and wahoo.
\newblock {\em Nature}, 202(4928):208, 1964.

\bibitem{bonner2015size}
John~Tyler Bonner.
\newblock {\em Size and cycle: an essay on the structure of biology}, volume
  2087.
\newblock Princeton University Press, 2015.

\bibitem{gough2021scaling}
William~T Gough, Hayden~J Smith, Matthew~S Savoca, Max~F Czapanskiy, Frank~E
  Fish, Jean Potvin, KC~Bierlich, David~E Cade, Jacopo Di~Clemente, John
  Kennedy, et~al.
\newblock Scaling of oscillatory kinematics and froude efficiency in baleen
  whales.
\newblock {\em Journal of Experimental Biology}, 224(13):jeb237586, 2021.

\bibitem{aoki2007measurement}
Kagari Aoki, Masao Amano, Naoki Sugiyama, Hiroyuki Muramoto, Michihiko Suzuki,
  Motoi Yoshioka, Kyoichi Mori, Daisuke Tokuda, and Nobuyuki Miyazaki.
\newblock Measurement of swimming speed in sperm whales.
\newblock In {\em 2007 Symposium on Underwater Technology and Workshop on
  Scientific Use of Submarine Cables and Related Technologies}, pages 467--471.
  IEEE, 2007.

\bibitem{lebastard_reactive_2016}
Vincent Lebastard, Frédéric Boyer, and Sylvain Lanneau.
\newblock Reactive underwater object inspection based on artificial electric
  sense.
\newblock {\em Bioinspir. Biomim.}, 11(4):045003, 2016.

\bibitem{zhu_tuna_2019}
J.~Zhu, C.~White, D.~K. Wainwright, V.~Di~Santo, G.~V. Lauder, and
  H.~Bart-Smith.
\newblock Tuna robotics: {A} high-frequency experimental platform exploring the
  performance space of swimming fishes.
\newblock {\em Science Robotics}, 4(34):eaax4615, 2019.

\bibitem{sanchez-rodriguez_proprioceptive_2021}
J.~Sánchez-Rodríguez, F.~Celestini, C.~Raufaste, and M.~Argentina.
\newblock Proprioceptive {Mechanism} for {Bioinspired} {Fish} {Swimming}.
\newblock {\em Phys. Rev. Lett.}, 126(23):234501, 2021.

\bibitem{thandiackal_emergence_2021}
Robin Thandiackal, Kamilo Melo, Laura Paez, Johann Herault, Takeshi Kano,
  Kyoichi Akiyama, Frédéric Boyer, Dimitri Ryczko, Akio Ishiguro, and Auke~J.
  Ijspeert.
\newblock Emergence of robust self-organized undulatory swimming based on local
  hydrodynamic force sensing.
\newblock {\em Science Robotics}, 6(57):eabf6354, August 2021.

\bibitem{lee_autonomously_2022}
Keel~Yong Lee, Sung-Jin Park, David~G. Matthews, Sean~L. Kim, Carlos~Antonio
  Marquez, John~F. Zimmerman, Herdeline Ann~M. Ardoña, Andre~G. Kleber,
  George~V. Lauder, and Kevin~Kit Parker.
\newblock An autonomously swimming biohybrid fish designed with human cardiac
  biophysics.
\newblock {\em Science}, 375(6581):639--647, 2022.

\bibitem{li2021burst}
Gen Li, Intesaaf Ashraf, Bill Fran{\c{c}}ois, Dmitry Kolomenskiy,
  Fr{\'e}d{\'e}ric Lechenault, Ramiro Godoy-Diana, and Benjamin Thiria.
\newblock Burst-and-coast swimmers optimize gait by adapting unique intrinsic
  cycle.
\newblock {\em Communications biology}, 4(1):1--7, 2021.

\bibitem{marras_fish_2015}
Stefano Marras, Shaun~S. Killen, Jan Lindström, David~J. McKenzie, John~F.
  Steffensen, and Paolo Domenici.
\newblock Fish swimming in schools save energy regardless of their spatial
  position.
\newblock {\em Behav Ecol Sociobiol}, 69(2):219--226, 2015.

\bibitem{wassersug1985kinematics}
RICHARD~J WASSERSUG and KARIN VON~SECHENDORF HOFF.
\newblock The kinematics of swimming in anuran larvae.
\newblock {\em Journal of experimental Biology}, 119(1):1--30, 1985.

\bibitem{webb_kinematics_1986}
Paul~W Webb.
\newblock Kinematics of lake sturgeon, {A}cipenser fulvescens, at cruising
  speeds.
\newblock {\em Canadian Journal of Zoology}, 64(10):2137--2141, 1986.

\bibitem{webb_steady_nodate}
Paul~W Webb.
\newblock Steady swimming kinematics of tiger musky, an esociform accelerator,
  and rainbow trout, a generalist cruiser.
\newblock {\em Journal of Experimental Biology}, 138(1):51--69, 1988.

\bibitem{rosenberger_functional_nodate}
Lisa~J Rosenberger and Mark~W Westneat.
\newblock Functional morphology of undulatory pectoral fin locomotion in the
  stingray {T}aeniura lymma ({C}hondrichthyes: {D}asyatidae).
\newblock {\em Journal of Experimental Biology}, 202(24):3523--3539, 1999.

\bibitem{magnuson1966courtship}
John~J Magnuson and John~H Prescott.
\newblock Courtship, locomotion, feeding, and miscellaneous behaviour of
  pacific bonito (sarda chiliensis).
\newblock {\em Animal behaviour}, 14(1):54--67, 1966.

\bibitem{mueller2010tail}
Anna-Maria Mueller, Deborah~L Burwen, Kevin~M Boswell, and Tim Mulligan.
\newblock Tail-beat patterns in dual-frequency identification sonar echograms
  and their potential use for species identification and bioenergetics studies.
\newblock {\em Transactions of the American Fisheries Society},
  139(3):900--910, 2010.

\bibitem{thiem2015accelerometer}
JD~Thiem, JW~Dawson, AC~Gleiss, EG~Martins, A~Haro, T~Castro-Santos,
  AJ~Danylchuk, RP~Wilson, and SJ~Cooke.
\newblock Accelerometer-derived activity correlates with volitional swimming
  speed in lake sturgeon (acipenser fulvescens).
\newblock {\em Canadian Journal of Zoology}, 93(8):645--654, 2015.

\bibitem{webb1982swimming}
PW~Webb and Raymond~S Keyes.
\newblock Swimming kinematics of sharks.
\newblock {\em Fishery Bulletin}, 80(4):803--812, 1982.

\bibitem{graham1990aspects}
Jeffrey~B Graham, Heidi Dewar, NC~Lai, William~R Lowell, and Steve~M Arce.
\newblock Aspects of shark swimming performance determined using a large water
  tunnel.
\newblock {\em Journal of Experimental Biology}, 151(1):175--192, 1990.

\bibitem{yan2013interspecific}
G-J Yan, X-K He, Z-D Cao, and S-J Fu.
\newblock An interspecific comparison between morphology and swimming
  performance in cyprinids.
\newblock {\em Journal of Evolutionary Biology}, 26(8):1802--1815, 2013.

\bibitem{malik2020ontogeny}
Arif Malik, Kathryn~A Dickson, Takashi Kitagawa, Ko~Fujioka, Ethan~E Estess,
  Charles Farwell, Kristy Forsgren, Jeannette Bush, and Kathryn~A Schuller.
\newblock Ontogeny of regional endothermy in pacific bluefin tuna (thunnus
  orientalis).
\newblock {\em Marine Biology}, 167(9):1--20, 2020.

\bibitem{dissanayake2008fishery}
DCT Dissanayake, EKV Samaraweera, and C~Amarasiri.
\newblock Fishery and feeding habits of yellowfin tuna (thunnus albacares)
  targeted by coastal tuna longlining in the north western and north eastern
  coasts of sri lanka.
\newblock {\em Sri Lanka J Aquat Sci}, 13:1--21, 2008.

\bibitem{mehanna2021length}
Sahar~F Mehanna and Alam~Eldeen Farouk.
\newblock Length-weight relationship of 60 fish species from the eastern
  mediterranean sea, egypt (gfcm-gsa 26).
\newblock {\em Frontiers in Marine Science}, page 942, 2021.

\bibitem{motta2013technical}
FS~Motta, FP~Caltabellotta, RC~Namora, and OBF Gadig.
\newblock Technical contribution length-weight relationships of sharks caught
  by artisanal fisheries from southeastern brazil.
\newblock {\em J. Appl. Ichthyol}, 1:2, 2013.

\bibitem{fish1984kinematics}
Frank~E Fish.
\newblock Kinematics of undulatory swimming in the american alligator.
\newblock {\em Copeia}, pages 839--843, 1984.

\bibitem{clark1979kinematics}
Brian~D Clark and Willy Bemis.
\newblock Kinematics of swimming of penguins at the detroit zoo.
\newblock {\em Journal of Zoology}, 188(3):411--428, 1979.

\bibitem{kojeszewski2007swimming}
Tricia Kojeszewski and Frank~E Fish.
\newblock Swimming kinematics of the florida manatee (trichechus manatus
  latirostris): hydrodynamic analysis of an undulatory mammalian swimmer.
\newblock {\em Journal of Experimental Biology}, 210(14):2411--2418, 2007.

\bibitem{goldbogen2006kinematics}
Jeremy~A Goldbogen, John Calambokidis, Robert~E Shadwick, Erin~M Oleson, Mark~A
  McDonald, and John~A Hildebrand.
\newblock Kinematics of foraging dives and lunge-feeding in fin whales.
\newblock {\em Journal of Experimental Biology}, 209(7):1231--1244, 2006.

\bibitem{fish_kinematics_nodate}
Frank~E Fish, S~Innes, and K~Ronald.
\newblock Kinematics and estimated thrust production of swimming harp and
  ringed seals.
\newblock {\em Journal of Experimental Biology}, 137(1):157--173, 1988.

\bibitem{nishiwaki1950body}
Masaharu Nishiwaki.
\newblock On the body weight of whales.
\newblock {\em Scientific Reports of the Whales Research Institute},
  4:184--209, 1950.

\bibitem{robeck2005reproduction}
Todd~R Robeck, Steven~L Monfort, Paul~P Calle, J~Lawrence Dunn, Eric Jensen,
  Jeffrey~R Boehm, Skip Young, and Steven~T Clark.
\newblock Reproduction, growth and development in captive beluga
  (delphinapterus leucas).
\newblock {\em Zoo Biology: Published in affiliation with the American Zoo and
  Aquarium Association}, 24(1):29--49, 2005.

\bibitem{christiansen2019estimating}
Fredrik Christiansen, Mariano Sironi, Michael~J Moore, Mat{\'\i}as Di~Martino,
  Marcos Ricciardi, Hunter~A Warick, Duncan~J Irschick, Robert Gutierrez, and
  Marcela~M Uhart.
\newblock Estimating body mass of free-living whales using aerial
  photogrammetry and 3d volumetrics.
\newblock {\em Methods in Ecology and Evolution}, 10(12):2034--2044, 2019.

\bibitem{economos1983elastic}
AC~Economos.
\newblock Elastic and/or geometric similarity in mammalian design?
\newblock {\em Journal of Theoretical Biology}, 103(1):167--172, 1983.

\bibitem{webPageSpeedAnimals}
\url{https://www.speedofanimals.com}.

\bibitem{wikipediaMakoShark}
{Wikipedia contributors}.
\newblock Shortfin mako shark --- {Wikipedia}{,} the free encyclopedia.
\newblock
  \url{https://en.wikipedia.org/w/index.php?title=Shortfin_mako_shark&oldid=1118544271},
  2022.

\bibitem{webPageOrcinusOrca}
\url{www.livescience.com/32772-what-animal-is-the-fastest-swimmer.html}.

\bibitem{webPageRingedSeal}
\url{www.oceanwide-expeditions.com/to-do/wildlife/ringed-seal-1}.

\bibitem{webPageSeaLion}
\url{www.nationalgeographic.com/animals/mammals/c/california-sea-lion}.

\end{thebibliography}

\end{document}